\documentclass[aps,prr,reprint]{revtex4-2}

\usepackage{graphicx}
\usepackage{svg}
\usepackage{dcolumn}
\usepackage{bm}
\usepackage{braket}
\usepackage{amsmath}
\usepackage{amssymb}
\usepackage[mathcal]{eucal}
\DeclareMathAlphabet{\mathcal}{OMS}{cmsy}{m}{n}
\usepackage{hyperref}
\usepackage{url}
\usepackage[normalem]{ulem}

\hypersetup{hypertex=true,
colorlinks=true,
linkcolor=blue,
urlcolor = blue,
citecolor=blue}

\usepackage{color}

\begin{document}
\let\origsaddcontentsline\addcontentsline
\renewcommand{\addcontentsline}[3]{}

\title{Fractal deconfinement and confinement in Sierpinski ice}

\author{James Walkling}
\email{jamwalk@pks.mpg.de}
\author{Roderich Moessner}%
\affiliation{%
 Max Planck Institute for the Physics of Complex Systems, Nöthnitzer Strasse 38, 01187 Dresden, Germany
}%


\begin{abstract}
We study the six-vertex model on the Sierpinski gasket, a four-coordinated hierarchical fractal with Hausdorff dimension $d_f=\log_23$. Given the importance of dimensionality for the long-wavelength behavior of such models, we specifically consider correlations and confinement as a function of the vertex weights, with the equal-weight point corresponding to the ice model. We calculate the partition function and correlators recursively to obtain a rich phase diagram hosting many different regimes. While the  ice model  shows entropic charge confinement, a particular four-vertex limit exhibits fractal deconfinement, with the string joining the deconfined charges itself a statistical fractal with fractal dimension $d_l=\log_2(5/2)\approx 1.3$. Finally, we propose a setup as an artificial spin ice to enable experimental study of the rich phenomenology of Sierpinski ice and its generalisations.

\end{abstract}

\maketitle
The six-vertex model is an exactly solvable constrained classical statistical mechanics model on the square lattice \cite{Lieb1967TwoDimensionalIceEntropy,Lieb1967FModelAntiferroelectric,Lieb1967SlaterKDPModel, Sutherland1967HydrogenBonded}. The Ising spins of an ice, or spin ice \cite{BramwellGingras}, model, $\sigma_i=\pm 1$, on a four-coordinated graph are defined on the edges of the graph, and can be visualized as arrows pointing into or out of a vertex. The structure of the model is associated with the constraint that each vertex must have two spins pointing in and two pointing out (ice rules \cite{BernalFowler1933}). This gives rise to six possible vertex configurations.

Since its formulation, a number of key insights have emerged from the six-vertex model. The phase diagram with respect to the Boltzmann weights hosts BKT transitions which were an early example of physics beyond Ising universality \cite{LiebWu1972FerroelectricModels, Berezinskii1972DestructionLongRangeOrderII, KosterlitzThouless1973OrderingMetastability}. Furthermore, there is an extended region where the model is described by a Coulomb phase with an effective emergent height model \cite{Henley2011}. The name Coulomb reflects the nature of the excitations out of the six-vertex configurations, which are charges subject to an emergent two-dimensional (logarithmic) Coulomb potential. Its generalization  to $3$D yields an  emergent $U(1)$ gauge field in pyrochlore spin ice \cite{Castelnovo2008}. In emergent gauge theories, string and charge dynamics are an active area of study in both theory and experiment \cite{Pollmann2025, GonzlezCuadra2025, borla2025stringbreaking21dmathbbz2, stringbreaking2025expproposal}.

\begin{figure*}
    \centering
    \includegraphics[width=\linewidth]{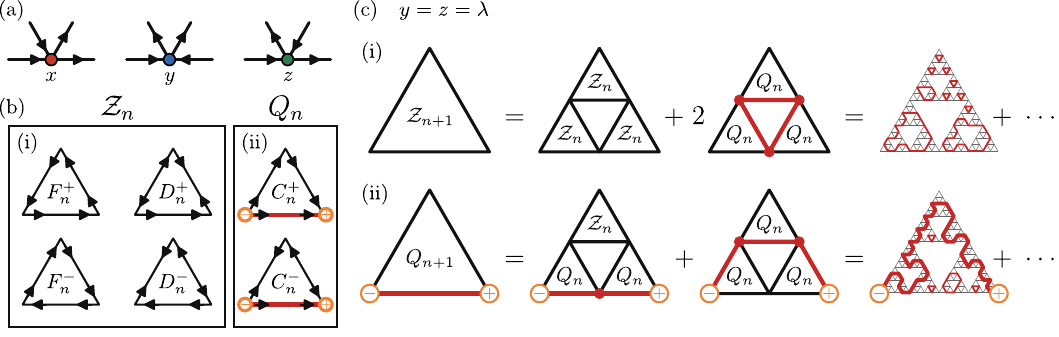}
    \caption{(a) three vertex configurations  not related by $\sigma_j \to -\sigma_j$ for the Sierpinski gasket six-vertex model. We color the vertices according to their type. They have Boltzmann weights $x,y,z$ such that $\alpha=\exp(-\beta E_\alpha)$ for $\alpha\in \{x,y,z\}$ with energies $E_\alpha$. (b) different recursion variables defined by their corner vertex orientations. (i) neutral vertices sum to $\mathcal{Z}_n$. (ii) two charge configurations sum to $Q_n$, with a red string joining the orange charges. (c) diagrams for the recursion relations of Eq.~\ref{eqn:ZQ_recur} where plaquettes are labeled based on their charge as in panel (b) and uncolored bulk vertices correspond to $y=z=\lambda$. (i) shows the recursion for $\mathcal{Z}_n$ where the factor of $2$ comes from the two possible directions of the loop. (ii) shows the recursion for $Q_n$, where the charges in orange are joined by a red string of $x$-type vertices.}
    \label{fig:introfig}
\end{figure*}

Fractal geometries offer an interesting yet tractable setting for many-body models, since their dilation symmetry can render them more solvable while often offering new phenomenology. Fractals in physics have a long history in phase transitions \cite{Mandelbrot1980, Suzuki1983PhaseTransitionFractals, Mandelbrot1984,Tarancon1998} and random walks \cite{RammalToulouse1983, Rammal1984, RW2003, RW2010} alongside current interest in numerous theoretical studies \cite{HaldaneSierpinski2024, LaughlinSierpinski2022, HofstadterSierpinski2018, QTransportSierpinski2024, Salvati2026NonErgodicMultifractalFractals, ToricCodeSG2022, biswas2026spin1diracdispersionchern} motivated by experimental realizations \cite{RydbergSierpinski2022, Zhou2022SGSpinWaves, Mehta2023SpinWaves, DaiYY2019SG, PhotonicTop2022, Li2023FractalPhotonicFloquet}. The defining behavior of fractals is scaling self-similarity, which occurs in two forms: deterministic fractals with hierarchical structure generated recursively according to a set rule (such as the Sierpinski gasket in Fig.~\ref{fig:introfig}(c)), and stochastic fractals which are only statistically self-similar (such as random walks in spin ice \cite{BSM2022}).

The prototypical example of a deterministic fractal, the Sierpinski gasket, has a fractal dimension $d_f=\log_23\approx 1.58$ describing the scaling of the number of edges, $N_l$, with the side length, $L$, such that $N_l=L^{d_f}$. Alongside various fractal dimensions, the ramification also plays an important role; the Sierpinski gasket has a finite ramification, since to disconnect $\mathcal{O}(L^{d_f})$ edges from the graph, only $\mathcal{O}(1)$ cuts are necessary \cite{Kirillov2013}. This strongly influences its properties; for finite ramification, the Ising model does not order at finite temperature \cite{Mandelbrot1984}. 

Here, we study the Sierpinski six-vertex model. This is a natural setting, since every vertex is four-coordinated (with the exception of the three corner vertices). In contrast to the square lattice, finite ramification and fractality of the Sierpinski gasket lead to a new phenomenology. We generalize previous work to generate coupled recursion relations for the partition function \cite{CHANG2013}. We thus show analytically that the charges at the degenerate ice point are entropically confined. 
However,  the limit of a four-vertex model exhibits 
a vanishing string tension: over an extended deconfined region, the string joining the charges fluctuates hierarchically and strongly, scaling with subfractal dimension $d_l=\log_2(5/2) < d_f$. 

Finally, we show how Sierpinski spin ice, and planar ice models generally, may be realised experimentally as an artificial spin ice (ASI), resolving a long-standing experimental obstacle in this field.

\textit{Sierpinski six-vertex model---} The six 
reduce to three unique vertices under Ising symmetry $\sigma_j\to -\sigma_j$ (Fig.~\ref{fig:introfig}(a)). They are parameterized by the Boltzmann weights $(x,y,z)$, and referred to as $x,y,z$-type vertices. Single spin flips are forbidden in the six-vertex model, since they give rise to violations of the ice rule. However, flipping directed loops of spins creates no charges, and hence maps directly between six-vertex states. The directedness of the loops ensures that the entry and exit edges  
($\sigma_\text{in}=-\sigma_\text{out}$) ensures preservation of the ice rules.

Violations of the six-vertex rule 
are excitations of the model, with charges $\pm2,\pm4$ depending on the imbalance of spins pointing into or out of a vertex. The eight total vertex configurations with three equal spins (charge $\pm2$) are taken to be degenerate and much higher in energy than the six-vertex states. There are also two excitations which have four equal spins (charge $\pm 4$), but these are taken to have a far higher energy and are neglected.

Depending on model parameters, charges are either deconfined or confined. In the latter case, they have a potential $\beta V(r)=\tau r$ with non-zero string tension $\tau$ for {Euclidean} distance $r$.

The number of generations of the hierarchical Sierpinski gasket is denoted by $n$, with $n=0$ is a single triangle, and $n\to \infty$ a self-similar fractal. The $n$th generation is composed of three copies of the $(n-1)$st generation recursively. The side length $L=2^n$, and there are $N_v=3(3^n-1)/2$ bulk four-coordinated vertices alongside the $3$ two-coordinated corner vertices.

\textit{Recursion Relations---}The hierarchical structure yields recursion relations for the partition function, $\mathcal{Z}_n$, calculated to $n>30$ with little effort. 
Recursion relations were previously derived for $x=y=z=1$ to study the  ice entropy  \cite{CHANG2013}; we generalize the equations to cover the full six-vertex model and to calculate observables.

The case, $y=z\equiv \lambda$ for general $x$, requires only two variables for the recursion relations: $\mathcal{Z}_n$, the charge-free partition function (Fig.~\ref{fig:introfig}(b)(i)), and $Q_n$, the constrained partition function for configurations with a pair of opposite charges on corner vertices (Fig.~\ref{fig:introfig}(b)(ii)). The recursion relations come from considering all combinations of joining $\mathcal{Z}_n, Q_n$ such that there are no charges in the bulk ($\mathcal{Z}_{n+1}$), or there are opposite charges on the corner vertices ($Q_{n+1}$). 
Directly reading from Fig.~\ref{fig:introfig}(c), we find:
\begin{equation}
    \begin{aligned}
            \mathcal{Z}_{n+1} &=\lambda^3\mathcal{Z}_n^3+2x^3Q_n^3,\\Q_{n+1}&=\lambda x^2Q_{n}^3 + \lambda^2 x Q_n^2\mathcal{Z}_n.
    \end{aligned}
    \label{eqn:ZQ_recur}
\end{equation}
The full recursion relations are included in the End Matter. Briefly: for $y\neq z$, $\mathcal{Z}_n$ must be written as separate variables, $F^{\pm}_n$ and $D^{\pm}_n$, which distinguish the two possible chiralities of the local spin circulation at neutral corner vertices as in Fig.~\ref{fig:introfig}(b)(i). For general $(x,y,z)$, we have  six recursion variables, Fig.~\ref{fig:introfig}(b). If the spin flip symmetry $\sigma_j \to -\sigma_j$ is unbroken, $F_n^\pm=F_n$ and $D_n^\pm=D_n$ but not necessarily $C^+_n\neq C^-_n$, Fig.~\ref{fig:introfig}(b)(ii).  

Defining $Q_n=C_n^++C_n^-$ and $\mathcal{Z}_n=F_n^++F_n^-+3(D_n^++D_n^-)=2F_n+6D_n$, the key parameters are
\begin{align}
P_n
&= D_n/F_n.
\label{eq:Pn_def}
\\
R_n
&= \mathcal{Z}_n/Q_n,
\label{eq:Rn_def}
\end{align}

$P_n$ is a measure of spin correlations, since $D_n$ and $F_n$ are defined by the chirality of the corner vertices (see Fig.~\ref{fig:introfig}(b)(i)). $P_n=1$ describes paramagnetic-like behavior where the three corner vertex chiralities are independent. The other extreme $P_n=0$ corresponds to $D_n=0$ such that the chiralities on corner vertices are ordered.
$R_n$ in turn measures the cost of a string: it gives the relative weight of a fully charge neutral system ($\mathcal{Z}_n$) and one with a string connecting two opposite charges on a pair of corner vertices ($Q_n$).

\begin{figure}
    \centering
    \includegraphics[width=\linewidth]{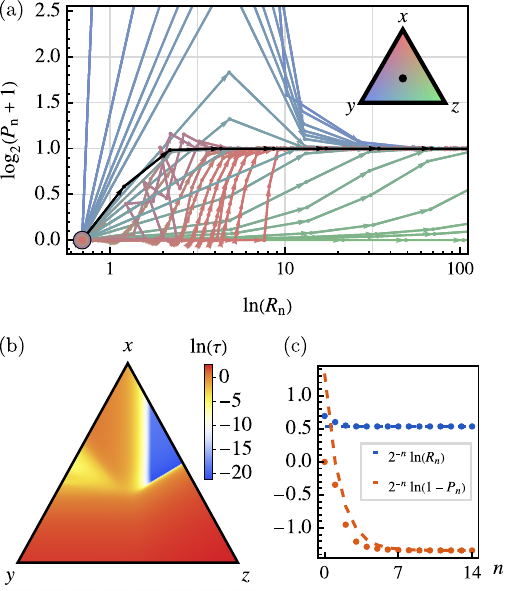}
    \caption{$P_n$ and $R_n$, Eqs.~\ref{eq:Pn_def},\ref{eq:Rn_def}, for different $(x,y,z)$. (a) parameter flow with $n$ projected onto recursion variables $P_n$ and $R_n$. The black line is for  ice $x=y=z$. For sufficiently large $n$, all phases with finite $x,y,z$ flow to a confining point with $P_*=1$, $R_n\to \infty$. There is an unstable fixed point with respect to $y,z>0$ marked by the circle in the lower left. (b) logarithm of the string tension, $\ln(\tau)=\ln(2^{-n}\ln R_n)$, for $n=30$. We plot a section of the space of $(x,y,z)$ by fixing $xyz=1$ where the parameters in the lower left corner are $(x,y,z)=(e^{-5},e^{-5},e^{10})$. The finite value of the tension indicates  confinement almost everywhere, excluding two disjoint regions with low tension. (c) asymptotic solutions of $R_n$ and $1-P_n$ (dashed lines) with their exact values (points) for the  ice case.}
    \label{fig:fig2}
\end{figure}

Fig.~\ref{fig:fig2}(a) shows the flow of $P_n, R_n$ with increasing $n$ for several $(x,y,z)$. Both $P_*=1$ and $P_*=0$ are fixed points, but only $P_*=1$ is stable to $y,z>0$. Even when microscopic couplings strongly break degeneracy, e.g.\ $y\gg z$, upon coarse-graining, a disordered mix of vertices emerges. Furthermore, $R_*=\infty$ is a fixed point of the recursion for several parameters, such that the shortest string has minimimum free energy.

\textit{Sierpinski ice---}
For $y=z\equiv \lambda$, Eq.~\ref{eq:Rn_def} yields
\begin{equation}
    R_{n+1}=\frac{\lambda^3 R_n^3+2x^3}{\lambda^2 xR_n+\lambda x^2}.
    \label{eqn:Ry=z_recur}
\end{equation}
For the ice point, $x=\lambda$, the asymptotic behavior is 
\begin{equation}
    R_n\sim \alpha^{2^n}=\alpha^{L},
    \label{eqn:asym_iso}
\end{equation}
where $\alpha\approx 1.7$. For different values of $x/\lambda$, the value $\alpha>1$ varies, but $\lim_{n\to \infty} R_n\to \infty$ regardless of the value of $x$. Confinement is also  demonstrated by the finite string tension, $\tau$, in Fig.~\ref{fig:fig2}(b). In Fig.~\ref{fig:fig2}(c), the asymptotic form given by Eq.~\ref{eqn:asym_iso} (dashed line), is compared with the full recursion data (points), showing convergence with growing $n$. The asymptotic result for $P_n$ is calculated in the Supplementary Material. We now calculate a number of properties, whose corresponding formulae are derived in the End Matter diagrammatically.

The degeneracy, and hence entropy, of the ice point $(x=y=z=1)$ is found via Eq.~\ref{eqn:ZQ_recur} as $\mathcal{Z}_{n+1}=\mathcal{Z}^3_n(1+2R_n^{-3})=\mathcal{Z}_n^3+\mathcal{O}(R_n^{-3})$ \cite{CHANG2013}. From Eq.~\ref{eqn:asym_iso}, the $R_n^{-3}$ correction vanishes superexponentially with $n$. 
We find $\mathcal{Z}_n\sim\Omega^{3^n}$ such that
\begin{equation}
    S_0/k_B=\ln(\Omega_n)\sim3^{n}\ln(\Omega)\sim L^{d_f}
    \label{eq:GSDegen}
\end{equation} is extensive, $N_l$, 
reflecting the fractal dimension $d_f=\log_2(3)$ of the Sierpinski gasket.

As discussed earlier, $P_n$ determines the connected correlator between spins on corner vertices. 
For any  $x,y,z$, 
\begin{equation}
    \langle \sigma_0 \sigma_{L} \rangle =\frac{1-P_n}{1+3P_n}.
    \label{eqn:corr}
\end{equation}
The asymptotic solution at the ice point gives the form $\langle \sigma_0 \sigma_{L} \rangle_{x=y=z}\sim 2 \alpha^4\Omega^{-2} \exp[-5\ln(\alpha)L/2],$ with $\Omega$ defined from Eq.~\ref{eq:GSDegen}. Correlations decay to zero exponentially as $P_n\to 1$ with a short correlation length $\xi=2/(5\ln(\alpha))\approx 0.75$ in units of the edge length.

Beyond the six-vertex configurations, $R_n$ describes charges and their connecting strings. The effective potential, $\beta V(L)$ associated with separating a pair of charges a distance $L$ along the bottom of the Sierpinski gasket is
\begin{equation}
    \beta V(L=2^n)=-\ln(Q_n)+\ln(\mathcal{Z}_n)=\ln(R_n),
    \label{eqn:Df}
\end{equation}
a difference in free energies between a chargeless and charged pair of corners. Asymptotically $\beta V(L) =L\ln(\alpha)$, defining  a string tension, $\tau=\ln\alpha$ between the charges, and hence entropic confinement. By contrast, confinement in square six-vertex models, as in an F-type model, typically arises only when breaking the  degeneracy between the vertices, so that strings already acquire a `bare' tension \cite{Lieb1967FModelAntiferroelectric, SpinIceKasteleyn2008}. In either case, the string predominantly stays close to the shortest path between the two charges.
The confinement potential derived above approximately generalizes to charges in the bulk. The subtle caveat is that the length $r$ is the shortest number of edges between the charges rather than their Euclidean distance. In the End Matter, we show the confinement result still asymptotically holds for separation $L/2$, and we derive $1/R_n$ corrections in the Suppl. Mat.

Since the fluctuations are described by $R_n$, this determines the fractal dimension, $d_l$, of the string averaged over all states contributing to $Q_n$, $\langle l \rangle_n=2^{d_l}\langle l\rangle_{n-1}$:
\begin{equation}
     d_l=\log_2 \bigg(\frac{2R_{n-1}+3}{R_{n-1}+1} \bigg),
     \label{eq:dl}
\end{equation}
which gives $d_l=1$ from Eq.~\ref{eqn:asym_iso}. To avoid confinement and $1$D scaling of the string, $R_n$ must be finite as $n\to\infty$ .

Confinement thus emerges fundamentally from  the finite ramification restricting  correlations between spins. However, this can be overcome by breaking vertex degeneracy. Fig.~\ref{fig:fig2}(b) shows two regions with finite $R_n$ such that the string tension vanishes, $\tau\to 0$: the point $x=y,z=0$ and the line $x>z,y=0$. We focus on the latter regime and derive its fractal $d_l=\log_2(5/2)$.

\textit{Fractal Deconfinement---}
\begin{figure}
    \centering
    \includegraphics[width=\linewidth]{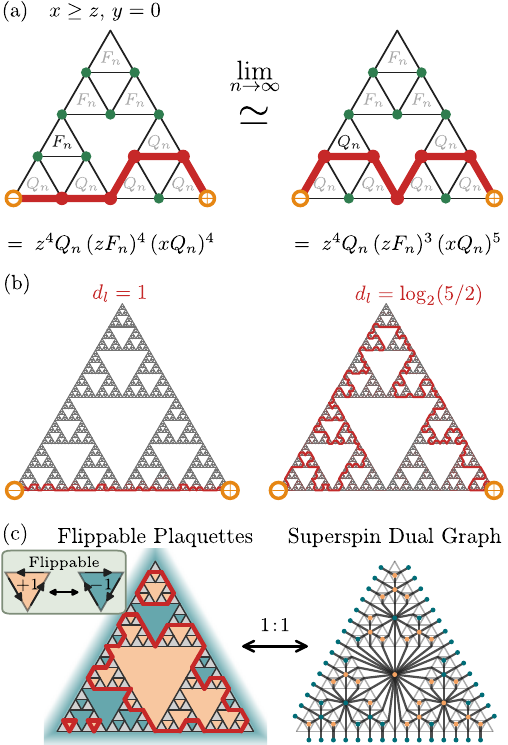}
    \caption{Fractal deconfinement and superspins. (a) Two contributions to $Q_{n+2}$, with gray recursion variables indicating unchanged plaquettes. Below the diagrams are their corresponding weights with ratio is $xQ_n/zF_n=1$ in the limit $n\to\infty$ for $x\geq z,y=0$. (b) Typical string joining two charges with different fractal dimensions, $d_l$ (Eq.~\ref{eq:dl}) for $n=6$ generations. (c) Flippable dual plaquettes and superspin construction for the $y=0$ four-vertex model overlayed on the fractal. Highlighted dual plaquettes have boundaries of flippable loops of spins, and different domains are separated by strings of $x$-type vertices. The $p$th dual plaquette can be treated as an Ising superspin with possible values $S_p=\pm 1$ depending on the circulation around its boundary. The upward-pointing gray triangles are not flippable. All edges out of the superspin dual graph connect to the same superspin.}
    \label{fig:fig3}
\end{figure}
Within the four-vertex model with $y=0$, there is a fractally deconfined regime for $x\geq z$. Alongside energetically favoring long strings and loops ($x>z$), setting $y=0$ reduces the number of accessible configurations.
The fixed point has a finite value $R_*=2x/z$ (see End Matter), which gives 
\begin{equation}
    F_n/Q_n\sim x/z
    \label{eqn:asymsolhAFM}
\end{equation} for large $n$. If we interpret $F_n,Q_n$ as some number of configurations associated with the internal structure after coarse-graining, Eq.~\ref{eqn:asymsolhAFM} implies that these balance the Boltzmann weights. Thus, the free energy difference between any pair of string configurations is zero, such that $\tau=0$. We show an illustrative example in Fig.~\ref{fig:fig3}(a); each segment of string added replaces $F_n\to Q_n$ and $z\to x$, such that as $n\to\infty$ the ratio of the weights of the diagrams is $(zF_n/xQ_n)^{\Delta l}=1$ where $\Delta l=1$ is the number of string segments they differ by.

The tensionless string with finite $R_*=2x/z$ gives rise to a non-trivial fractal string dimension, $d_l$. While Eq.~\ref{eq:dl} only applies to the case $y=z$, we can calculate $d_l$ directly for $\tau=0$. When comparing the $n$th generation to the $(n+1)$th generation in Fig.~\ref{fig:introfig}(c)(ii) for $Q_{n+1}$, there are two string configurations that contribute: one with length $l_{n+1}=2l_n$ and another with $l_{n+1}=3l_n$. All strings are equally weighted for $\tau=0$, so their mean length obeys $\langle l\rangle_{n+1}=\frac{5}{2}\langle l\rangle_n$. Hence, the scaling when we double $L\to 2L$ is $\langle l\rangle_n\to 2^{d_l}\langle l\rangle_n$ where $d_l=\log_2(5/2)$ describes a statistical fractal, Fig.~\ref{fig:fig3}(b).

This fractally deconfined regime
can be further understood via
an emergent Ising superspin model. The low-energy states are four-vertex configurations connected by flipping directed loops of spins. When $y=0$, the set of allowed directed loops is disjoint: any allowed configuration for $y=0$ is specified by the circulation ($S_p=\pm 1$) of each of the highlighted dual plaquettes $p$ in Fig.~\ref{fig:fig3}(c). The 
effective superspins $S_p=\pm 1$ thus encode all four-vertex ($y=0$) configurations. There are $N_S=(3^n+1)/2$ different values of $p$ for $S_p$. The superspins are hierarchically composed of $3\times 2^m$ spins $\sigma_j$ where $m=0,...,n$. $S_p,S_q$ couple via effective two-body interactions $\tilde{J}_{pq}=\frac{1}{2}\ln(x/z)$ per bond, since the relative superspins determine if their shared vertex is $x$ or $z$. For $x>z$, this is an effective antiferromagnet.

The superspin picture directly reveals the structure of the four-vertex model, analogously to the height model construction for the square lattice Coulomb phase \cite{Henley2011}. The number of low-energy four-vertex configurations is exactly $2^{N_s}$ from the $N_s$ independently flippable superspins, and the spin correlations are $\langle \sigma_i \sigma_j \rangle =1$ iff $i,j$ are part of the same superspin $S_p$. This is in agreement with Eq.~\ref{eqn:corr}, since $P_n=0$ when $y=0$ such that $\langle \sigma_0 \sigma_L\rangle=1$.

For charge excitations, the superspin construction breaks down, because superspins cease to have definite circulation. However, we can treat the charges as preventing a superspin from flipping, such that a pair of charges in $S_p$ reduces the entropy by $\Delta S=-k_BT\ln(2)$. When $x=z$, there is no energy difference, and $\tau=0$ arises purely from the lack of change in entropy when moving charges inside of $S_p$. The rich interplay between geometry, energy and entropy generalizes this $\tau=0$ result to all $x\geq z$. Hence, we find that the excitations of the model are restricted to disjoint quasi-$1$D loops, although the string is a delocalized statistical fractal.

The results above strictly hold for $L\to \infty$, and the convergence to $R_*=2x/z$ for large $L=2^n$ is
\begin{equation}
    R_n/R_*\sim1-\frac{1}{n}= 1 - \frac{1}{\log_2(L)}.
\end{equation}
Although the logarithmic convergence is slow, $R_0=2$ always, such that for $x\approx z$, we start very close to the fixed point with good finite-size approximation to $L\to \infty$. We derive the scaling in detail in the Suppl.\ Mat.

\begin{figure}
    \centering
    \includegraphics[width=\linewidth]{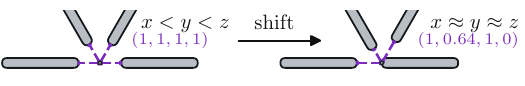}
    \caption{Potential experimental implementation of degenerate vertices in artificial spin ice (ASI). The distances of the tips of the micromagnetic islands  islands (gray) from the central vertex are shown in purple. They are tuned to the values $(1,0.64,1,0)$ such that $x\approx y\approx z$ as calculated within the dumbbell approximation \cite{Castelnovo2008, Moller2009MultipoleASI}.}
    \label{fig:fig4}
\end{figure}

\textit{Experimental Implementation---}%
We propose artificial spin ice (ASI) \cite{2013spinicereview, 2019spinicereview}, an array of nanomagnetic islands, as a promising experimental platform 
for probing fractal ice. The difficulty -- as already noted in the foundational experiment \cite{Wang2006} -- is that simply replacing edges with magnetic islands does not yield  degenerate vertex configurations, but rather a fixed choice of relative vertex weights. For square ASI, one possible resolution is the introduction of a height offset \cite{Moller2006ArtificialSquareIce, Perrin2016SquareIce}, while another deforms the square vertex into a rhombus \cite{Ribeiro2017RectangularASI, Nascimento2024GaugeFieldPlanarASI}. Here, we show that a general strategy for constructing tunable vertex weights is to vary the relative distances of the tips of the islands in the vertex, as illustrated in Fig.~\ref{fig:fig4}. 

We  concretely (and quantitatively) illustrate this idea within the dumbbell model of spin ice \cite{Castelnovo2008,Moller2009MultipoleASI}, which applies 
to the case of narrow islands of uniform width. 
Here, the dipolar island is represented by a pair of opposite charges at its endpoints matching its magnetic moment. Crucially, computing the Coulomb interaction between these charges as a function of their relative locations robustly yields a one-parameter family of {\it vertex-degenerate} configurations, of which one member is shown in Fig.~\ref{fig:fig4}. 

We emphasize that this scheme applies generally, for more or less exotic geometries, and it can be used to generate not only vertex-degenerate settings, but also, e.g., ones in the parameter regime of the fractal deconfined regime. More details are given in the Supplementary Material. Also, an  implementation beyond the dumbbell model, i.e.\ for more general island geometries, can use quantitative input from micromagnetic simulations.

\textit{Outlook---}Beyond the Sierpinski gasket, there are a number of other geometries of interest to which our techniques apply. Other fractals, such as the Sierpinski carpet, have infinite ramification which could lead to different correlations, in analogy to the Ising model \cite{Mandelbrot1984III}. Additionally, our techniques  can be extended to fractals embedded in $3$D. The Sierpinski pyramid is a particularly promising candidate to understand corrections due to loops on the pyrochlore lattice, going beyond the previously studied tree-structure of the Husimi cactus \cite{Castelnovo2008, HusimiCactus2018}.

The model also has possible extensions that would be well-suited to experimental platforms. ASI offers a rich platform in which to study annealing of the string in the fractal geometry and potential exotic fractal contributions to the dynamics using our proposal in Fig.~\ref{fig:fig4} \cite{2013spinicereview, 2019spinicereview}. Furthermore, NISQ platforms offer a setting to further explore quantum effects on the strings and charges. The behavior of charges and strings in quantum models is an active area of research to which this model could be generalized \cite{Pollmann2025,GonzlezCuadra2025,Rydberg_Stringbreaking2025,borla2025stringbreaking21dmathbbz2}.

\begin{acknowledgments}
RM thanks Flavio Seno and Julia Yeomans for an earlier collaboration on the Sierpinski gasket, and Peter Schiffer for discussions which led to this work.
This work was supported in part by the Deutsche Forschungsgemeinschaft via the cluster of excellence ctd.qmat (EXC 2147, project-id 390858490) and  FOR 5522 (Project-ID No. 499180199).
\end{acknowledgments}


\bibliography{apssamp}

\clearpage

\begin{widetext}
\begin{center}
\textbf{\large End Matter:\\[0.3em]
Fractal Scaling in the Six-Vertex Model on the Sierpinski Gasket}
\end{center}

\textit{Appendix A: Complete Recursion Relations---}The full recursion relations for general $(x,y,z)$ are given below. The definition of the corresponding recursion variables is that they count the weighting of the states for a given set of corner vertex configurations as shown in Fig.~\ref{fig:introfig}. For the neutral configurations we have
\begin{equation}
\begin{aligned}
F_{n+1} &=
(C_n^-)^3 x^3 + (C_n^+)^3 x^3
+ 2D_n^3 y^3 + 6D_n^2F_n y^3
+ 9D_n^3 y^2z + 12D_n^2F_n y^2z + 3D_nF_n^2 y^2z \\
&\quad
+ 12D_n^3 yz^2 + 6D_n^2F_n yz^2 + 6D_nF_n^2 yz^2 
+ 4D_n^3 z^3 + 3D_n^2F_n z^3 + F_n^3 z^3,
\end{aligned}
\end{equation}
and the relation
\begin{equation}
\begin{aligned}
D_{n+1} &=
(C_n^-)^2 C_n^+ x^3 + C_n^- (C_n^+)^2 x^3 
+ 4D_n^3 y^3 + 2D_n^2F_n y^3 + 2D_nF_n^2 y^3 \\
&\quad
+ 10D_n^3 y^2z + 11D_n^2F_n y^2z + 2D_nF_n^2 y^2z + F_n^3 y^2z \\
&\quad
+ 10D_n^3 yz^2 + 10D_n^2F_n yz^2 + 4D_nF_n^2 yz^2 
+ 3D_n^3 z^3 + 4D_n^2F_n z^3 + D_nF_n^2 z^3. 
\end{aligned}
\end{equation}
For the configurations with charges fixed on two of the three corner vertices, we find that
\begin{equation}
\begin{aligned}
C_{n+1}^+ &=
2C_n^- (C_n^+)^2 x^2y
+ 2C_n^- C_n^+ D_n xy^2
+ (C_n^+)^2 D_n xy^2
+ (C_n^-)^2 F_n xy^2
+ (C_n^-)^2 C_n^+ x^2z
+ (C_n^+)^3 x^2z \\
&\quad
+ 2(C_n^-)^2 D_n xyz
+ 2C_n^- C_n^+ D_n xyz
+ 2(C_n^+)^2 D_n xyz
+ 2C_n^- C_n^+ F_n xyz \\
&\quad
+ (C_n^-)^2 D_n xz^2
+ 2C_n^- C_n^+ D_n xz^2
+ (C_n^+)^2 F_n xz^2,
\end{aligned}
\end{equation}
and for the other chirality of the neutral vertex relative to the position of the charged vertices,
\begin{equation}
\begin{aligned}
C_{n+1}^- &=
2(C_n^-)^2 C_n^+ x^2y
+ (C_n^-)^2 D_n xy^2
+ 2C_n^- C_n^+ D_n xy^2
+ (C_n^+)^2 F_n xy^2
+ (C_n^-)^3 x^2z
+ C_n^- (C_n^+)^2 x^2z \\
&\quad
+ 2(C_n^-)^2 D_n xyz
+ 2C_n^- C_n^+ D_n xyz
+ 2(C_n^+)^2 D_n xyz
+ 2C_n^- C_n^+ F_n xyz \\
&\quad
+ 2C_n^- C_n^+ D_n xz^2
+ (C_n^+)^2 D_n xz^2
+ (C_n^-)^2 F_n xz^2. 
\end{aligned}
\end{equation}
These formulae have been simplified from the most general case, since $F^+_n=F_n^-=F_n$ and $D_n^+=D_n^-=D_n$ in the absence of external symmetry-breaking fields. The initial conditions are always $F_0^+=F_0^-=C_0^+=1$ with all other recursion variables set to $0$.

\textit{Appendix B: Recursion for Hierarchical Antiferromagnet---}
In the limit $y=0$, we find that $D_n^-=D_n^+=C_n^-=0$ for all $n$. This reduces the number of non-zero recursion variables to three:
\begin{equation}
\begin{aligned}
    F_{n+1}^+=z^3(F_n^+)^3+x^3 (C^+_n)^3,\quad F_{n+1}^-=z^3(F_n^-)^3+x^3 (C^+_n)^3,\quad C_{n+1}^+=xz^2 (F_n^-)(C_n^+)^2+x^2z(C_n^+)^3.
\end{aligned} 
\label{eqn:yforbidrecurse}
\end{equation}

We find that the ratio 
\begin{equation}
    R_{n+1}=\frac{\mathcal{Z}_{n+1} }{Q_{n+1}}=\frac{F_{n+1}^++F_{n+1}^-}{C_{n+1}^+}=\frac{zR_n^2}{2x}-R_n+\frac{2x}{z}.
\end{equation}
Since $D_n=0$, we have $P_n=0$. Studying this non-linear map, we find that it has exactly one marginal fixed point at $R_*=2x/z$ which is semi-stable; it is attractive from below, but repulsive from above. Since $R_0=2$ always, depending on the value of $x/z$ we either flow to the fixed point or blow up.
\begin{equation}
R_{n}\sim\begin{cases}
\frac{2x}{z}\bigg(1-\frac{1}{n}+\mathcal{O}\bigg(\frac{\log(n)}{n^2}\bigg)\bigg), & x>z \\
 \gamma(x/z)^{2^n}, & x<z \\
2, & x=z,
\end{cases}
\label{eqn:Rny=0}
\end{equation}
for a function $\gamma(t)>1$ determined by the asymptotics.

These results are in line with the expectations for the phases, since $x<z$ gives rise to a trivial ferromagnetic phase ($R_n \to \infty$). However, when $x>z$, $R_n\to 2x/z$ even as $n\to \infty$ which indicates potentially non-trivial fractal behavior in this part of the phase diagram.

\textit{Appendix C: Deriving Observables---}
Below are diagrammatic derivations for the observables in the main text for $x=y=z=1$. The free energy change, $\Delta f$, in extending the string can be calculated by comparing the weight of a charge at $L$ vs. $L/2$: $\Delta f= f(L)- f(L/2)=-\ln(Q_n^{(n)})+\ln(Q_n^{(n-1)})$ where we set $\beta=1/k_BT=1$. $Q_n^{(n)}=Q_n$ is the constrained partition function with two charges on the corner vertices, and $Q_n^{(n-1)}$ is the case with one charge at position $L/2$ (see Fig.~\ref{fig:endmatter}(a)). Expressing these recursion variables in terms of the $(n-1)$th generation, we find that
\begin{equation}
    f=\ln \bigg( \frac{Q_n^{(n-1)}}{Q_n^{(n)}} \bigg)=\ln \bigg( \frac{\mathcal{Z}_{n-1}^2 Q_{n-1}+Q_{n-1}^3}{\mathcal{Z}_{n-1} Q^2_{n-1}+Q_{n-1}^3} \bigg)=\ln \bigg( \frac{R_{n-1}^2+1}{R_{n-1}+1} \bigg).
\end{equation}

The fractal dimension of the string, $d_l$ (Eq.~\ref{eq:dl}) can be similarly calculated. The average length of the string scales recursively via $\langle l \rangle _n= p_3\cdot 3\langle l \rangle_{n-1}+p_2 \cdot 2\langle l\rangle_{n-1}$ where $p_i$ is the probability of a string with $i$ segments between the charges. Comparing $n$ and $n-1$, there are two possible paths (Fig.~\ref{fig:endmatter}(b)). Using the diagrams, it follows that
\begin{equation}
    \langle l\rangle_n=\frac{(2Q_{n-1}^2\mathcal{Z}_{n-1}+3Q_{n-1}^3)}{Q_{n-1}^2\mathcal{Z}_{n-1}+Q_{n-1}^3}\langle l\rangle_{n-1}= \frac{2R_{n-1}+3}{R_{n-1}+1}\langle l\rangle_{n-1},
\end{equation}
where the dimension is defined by $\langle l \rangle_n=2^{d_l} \langle l\rangle_{n-1}$. 

The correlations, $\langle \sigma_0\sigma_L\rangle$, are calculated by considering the neutral configurations. We show the eight total terms in Fig.~\ref{fig:endmatter}(c): four diagrams have spins with the same direction (green) giving a positive contribution with total weight $2(F_n+D_n)$, and four diagrams give a negative contributions (red) for spins with opposite signs, $4D_n$. Summing, we find that
\begin{equation}
    \langle \sigma_0\sigma_L\rangle = \frac{2(F_n+D_n)-4D_n}{2(F_n+3D_n)}=\frac{1-P_n}{1+3P_n},
\end{equation}
using $P_n=D_n/F_n$ as stated in the main text. In the Supplementary Material, we evaluate these simple diagrams for the string tension rigorously in the bulk with a more complete analysis.

\begin{figure}
    \centering
    \includegraphics[width=1.0\linewidth]{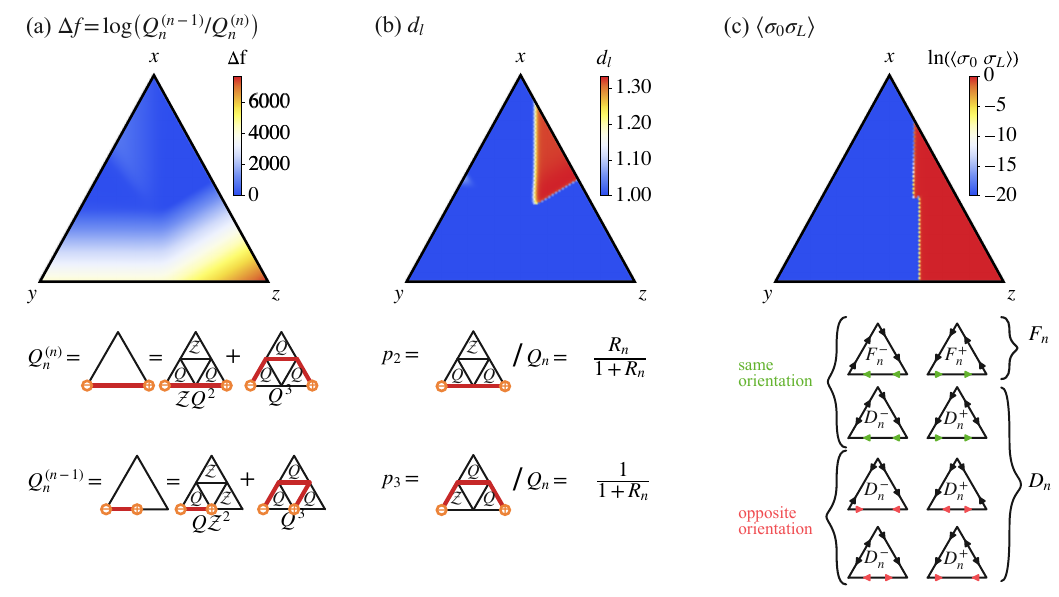}
    \caption{Illustrations of diagrams for calculating different observables of interest in the main text alongside full recursion calculations. (a) calculation of the string tension, $\Delta f$, which comes from studying the change in free energy as we move a charge from $L/2$ to $L$ along the bottom of the fractal. (b) fractal dimension, $d_l$, of the string and its definition via the lengths. The two possible paths at the largest length scales are 3 copies or 2 copies of the previous iteration with probabilities of $p_3$ and $p_2$ respectively. (c) neutral diagrams that contribute to the spin-spin correlator. The spins in color are those at $0$ and $L$. Same circulation configurations give a positive contribution to the correlator of $2(F_n+D_n)$, whereas opposite spins give a negative contribution $-4D_n$. }
    \label{fig:endmatter}
\end{figure}
\pagebreak

\end{widetext}

\end{document}


\title{Supplementary Material\\[0.3em]
Sierpinski Ice} 
\author{James Walkling}\affiliation{Max Planck Institute for the Physics of Complex Systems, Nöthnitzer Strasse 38, 01187 Dresden, Germany}
\author{Roderich Moessner}\affiliation{Max Planck Institute for the Physics of Complex Systems, Nöthnitzer Strasse 38, 01187 Dresden, Germany}

\maketitle

\tableofcontents

This supplementary material is organized as follows. In Sec.~\ref{sec:fourvert}, we discuss the four-vertex limits of the full six-vertex model in detail. In Sec.~\ref{sec:vertexorder} we study the vertex occupations of the full model over the full parameter space $(x,y,z)$ and demonstrate that four-vertex models play a key role in describing the low-energy physics in many different regimes, and name the different types of vertex order. Sec.~\ref{sec:asymrecur} gives comprehensive derivations for the asymptotic results across all the different four-vertex models and ice regime that support the results of the main text. In Sec.~\ref{sec:stringtensionbulk}, we demonstrate that the string tension results  for corner charges generalize to charges in the bulk in the two key different regimes (ice and HAFM). We conclude with Sec.~\ref{sec:ASIimplementation}, where we show that by adjusting the positions of the tips of the islands in artificial spin ice (ASI), one can tune the relative size of the Boltzmann weights $(x,y,z)$ of the model.

\section{Four-vertex Superspin Models}
\label{sec:fourvert}
The six-vertex model reduces to three different four-vertex models once one of $(x,y,z)$ is set to zero.

\begin{figure}
    \centering
    \includegraphics[width=\linewidth]{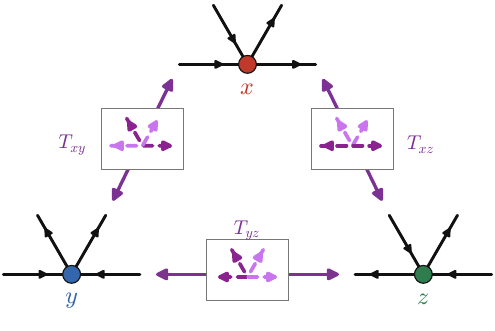}
    \caption{Transition paths (purple), $T_{\alpha \beta}$, between three different example vertices of $x,y,z$-type (red, blue, green) out of the six total vertices. The double arrows indicate that the paths can be walked in either direction to flip the spins. The darker purple dashed path indicates the transition path between the two example vertices shown, whereas the light purple is another possible transition path between two vertices of a given type.}
    \label{fig:pathtransitions}
\end{figure}

For the four-vertex model in the case of $y=0$ in the main text, we discussed how the allowed flippable loops that connect allowed configurations lead to an emergent superspin model. These flippable loops are paths on edges starting and ending at the same vertex that preserve the constraint at each vertex by exiting on an edge with opposite direction to that entered on. We give examples of possible choices of entry and exit edges for making transitions between different types of vertices as dashed lines in Fig.~\ref{fig:pathtransitions}. The dark purple path indicates the move taken to get between the two example configurations shown, and the light purple indicates another possible path between the given types of vertices. These single vertex moves are joined together to form the overall shape of the allowed loop. Once the loop is found, it is flipped by setting $\sigma_j \to -\sigma_j$ for the edges, $j$, in the loop. 

\begin{figure}
    \centering
    \includegraphics[width=\linewidth]{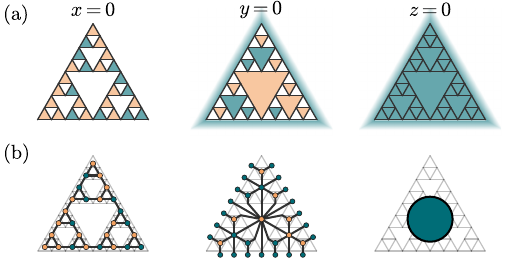}
    \caption{(a) flippable plaquettes colored in dark cyan ($S_p=-1$) and orange ($S_p=+1$). White plaquettes cannot be flipped without creating energetically forbidden vertices as specified by which Boltzmann weight is zero. The $x=0$ and $y=0$ flippable plaquettes are dual. There is only one flippable loop for $z=0$ (shown in Fig.~\ref{fig:eulercircuit}). (b) corresponding superspin graphs for the plaquette models in (a). Each black edge represents a coupling $\tilde{J}_{pq}$ between the pair of plaquettes as in Eq.~\ref{eqn:superspin}. All the vertices outside of the second figure for $y=0$ correspond to a single boundary superspin. There are no edges for the third graph of $z=0$, since there is only a single superspin.}
    \label{fig:superspins}
\end{figure}

In contrast to the gas of intersecting flippable loops for the six-vertex model, the four-vertex flippable loops are {\it disjoint} sets of spins. Hence, four-vertex configurations of the system are fully specified by specifying the circulation, $S_p=\pm 1$, of the areas enclosed by these sets of spins. Fig.~\ref{fig:superspins}(a) shows circulations of $S_p=+1$ (orange) and $S_p=-1$ (dark cyan).

By studying the connectivity of the plaquettes via shared vertices, the low-energy physics of the model (for no charges on vertices, $Q_v=0$), can be captured by a model with effective Ising spins $S_p=\pm 1$ as in Fig.~\ref{fig:superspins}(b). Each dark edge represents an interaction $\tilde{J}_{pq}$ corresponding to the effective Hamiltonian
\begin{equation}
    H=\sum_{i,j \in E} J_{ij} \sigma_i \sigma_j \xrightarrow{Q_v=0} H_0=\sum_{p,q \in \mathcal{P}} \tilde{J}_{pq} S_p S_q,
    \label{eqn:superspin}
\end{equation}
with $E$ defined as the set of edges and $\mathcal{P}$ as the set of plaquettes. We show explicitly how this claim holds as well in the limit of the other four-vertex models when $x=0$ or $z=0$ as shown in Fig.~\ref{fig:superspins}. Charges ($Q_v\neq 0$) can also be studied in the context of four-vertex models, although then the superspin construction must be applied more carefully. 

\subsection{$x=0$: Sierpinski Ising}
\label{sec:fourvertx=0}
In the limit of $x\to 0$, the superspins are defined as the $3^n$ smallest triangular plaquettes (Fig.~\ref{fig:superspins}(a) for $x=0$). If these plaquettes have the same (opposite) circulation, their shared vertex has weight $z$ ($y$) as shown in Fig.~\ref{fig:pathtransitions} for the move $T_{yz}$. Hence, for the couplings $\tilde{J}_{pq}$ between superspin plaquettes $p,q$, tuning $y$ and $z$ switches the model from an antiferromagnet (AFM, $y\gg z$) to a ferromagnet (FM, $z\gg y$). 
Although these effective spins live on the plaquettes, the thermodynamics is equivalent to the well-studied model of Ising spins on the vertices of the Sierpinski gasket due to the finite ramification of the leftmost graph in Fig.~\ref{fig:superspins}(b) \cite{Mandelbrot1980,Mandelbrot1984, Tarancon1998}. Much like the $1$D Ising model, this model hosts a paramagnet (PM) at any finite temperature, which corresponds to the instability of the FM point to perturbations $y>0$ in the projected RG flow of Fig.~2(a) from the main text. 

\subsection{$y=0$: Dual Hierarchical}
In the limit of $y\to 0$, the superspins define the circulation of the ``dual plaquettes" as shown in Fig.~\ref{fig:superspins}(a) for $y=0$. The corresponding superspins have a hierarchical ordering. At the smallest scales, three $\sigma_j$'s define a superspin $S_p$, and the largest superspin is the boundary of the fractal composed of $3\times 2^n$ $\sigma_j$. For the regime $x>z$, the superspins favour anti-alignment to create $x$-vertices on the shared vertex between the plaquettes, and for $x<z$, the plaquettes are energetically driven to align. While the latter case reduces to a FM, the case $x>z$ is an AFM living on the hierarchical graph in Fig.~\ref{fig:superspins}(b), hence the name hierarchical antiferromagnet (HAFM).

Since $T_{xy}$ moves are forbidden for $y=0$, the charge excitations of the model are restricted to a given superspin; moving an excitation from one superspin to another would require $T_{xy}$, which has an infinite energy barrier to produce a $y$-type vertex, or lead to the creation of a vertex with four arrows pointing in the same direction which is also energetically forbidden. Thus, the allowed motion of the charges in the $y=0$ four-vertex model is restricted to a set of disjoint quasi-$1$D loops defined by the superspins.

\subsection{$z=0$: Eulerian Circuit}
\label{sec:fourvertz=0}
\begin{figure}
    \centering
    \includegraphics[width=0.8\linewidth]{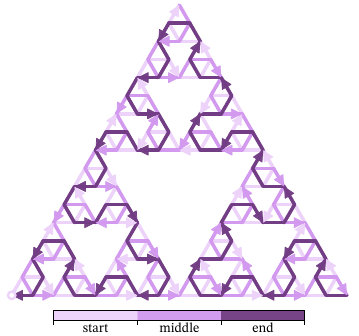}
    \caption{Joining together the allowed moves $T_{xy}$ defined in Fig.~\ref{fig:pathtransitions} covers every edge of the Sierpinski gasket exactly once as shown here for $n=4$. Note that this includes an allowed move to exit the three corners. This is an Eulerian circuit which here has been broken into three sequences of different brightness related by three-fold rotational symmetry.}
    \label{fig:eulercircuit}
\end{figure}

In the limit of $z\to 0$, the nature of the constraint only permits a single flippable loop, which follows from the shape of the path of the loop with only $T_{xy}$ in Fig.~\ref{fig:pathtransitions}. This loop is the unique Eulerian circuit: a path that covers every edge exactly once as shown in three segments in Fig.~\ref{fig:eulercircuit}. In the superspin construction, this means that the whole gasket forms a single superspin at any scale as shown in Fig.~\ref{fig:superspins}(a) for $z=0$. The single superspin picture agrees perfectly with the asymptotic calculations of Eq.~\ref{eqn:RnHFM}. 

If we assume that the charge excitations of the model have a much lower energy than the cost of $z$-type vertices, the motion is restricted to a fractal curve with dimension $d=\log_2 3$ (as shown in the darker shade of Fig.~\ref{fig:eulercircuit}). Despite the motion of the charges being confined to a fractal path, the string joining the charges scales trivially like $d_l=1$ in this regime.

\section{Vertex Orders}
\label{sec:vertexorder}
\begin{figure}
    \centering
    \includegraphics[width=\linewidth]{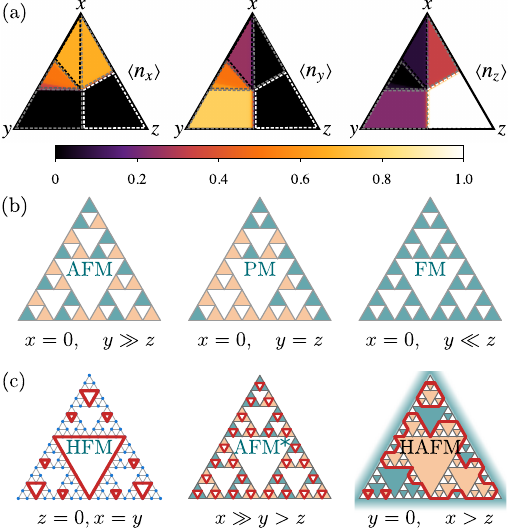}
    \caption{Typical states for the various different kinds of vertex order along their corresponding occupations of different types of vertices where the orange vs. dark cyan highlight indicates the circulation of the plaquette. (a) vertex occupations $\langle n_\alpha \rangle$ calculated for $n=20$ generations. The dashed lines highlight six-vertex (black), four-vertex (grey) and two-vertex (white) regimes in correspondence with Table.~\ref{tab:overviewofphases}. (b) vertex densities $\langle n_x \rangle, \langle n_y \rangle, \langle n_z \rangle$. (c) conventional vertex orders which are directly analogous to the AFM, PM and FM.}
    \label{fig:vertexorders}
\end{figure}

The system supports several distinct types of vertex order, characterized by the average occupation of each vertex type, $\langle n_\alpha\rangle=N_\alpha/N_v$, where $N_v=3(3^n-1)/2$ is the number of bulk vertices with four-fold coordination. Expressing this in terms of the partition function, $\mathcal{Z}_n$,
\begin{equation}
    \langle n_{\alpha}\rangle=\frac{1}{N_v}\frac{\partial \ln(\mathcal{Z}_n)}{\partial \ln(\alpha)}.
\end{equation} This is defined such that $0\leq \langle n_\alpha \rangle \leq 1$ and $\sum_\alpha \langle n_\alpha \rangle=1$. These occupations are shown in Tab.~\ref{tab:overviewofphases} and are plotted in Fig.~\ref{fig:vertexorders}. We derive their analytic values in detail, and see that the extremes of the model are just four-vertex models. The earlier discussed emergent superspin construction of Sec.~\ref{sec:fourvert} is a useful framework to describe their behavior.

$\langle \mathbf{n} \rangle=(\langle n_x \rangle, \langle n_y\rangle, \langle n_z \rangle)$ is sensitive to the microscopic structure: since there is no coarse-graining over the smallest degrees of freedom, every vertex is counted equally. This is in direct contrast to the $R_n$ and $P_n$ introduced in the main paper. A key example of the difference is that although a regime may have $\langle \mathbf{n} \rangle \approx (0,0,1)$ like a ferromagnet, it will still be a paramagnet in the coarse-grained sense, because as explained in the main text, new $y$ or $z$ vertices are equally favored by the free energy. Between these different vertex orders in Fig.~\ref{fig:vertexorders}(a) there are no thermodynamically sharp transitions, just crossovers, an example of which is shown in Fig.~\ref{fig:cascadecrossover}.

The vertex orders fall into two different categories: conventional order which can exist on any geometry, Fig.~\ref{fig:vertexorders}(b), and hierarchical order, Fig.~\ref{fig:vertexorders}(c), where the interplay between the strings of $x$-type vertices and hierarchical structure is constitutively related to the fractal nature of the Sierpinski gasket.  

\begin{table}[h!]
\centering
\small
\renewcommand{\arraystretch}{1.25}
\setlength{\tabcolsep}{4pt}

\begin{tabular}{p{0.35\linewidth} p{0.25\linewidth} p{0.3\linewidth}}
\hline
$(x,y,z)$
& \textbf{Vertex order}
& $(\langle n_x \rangle, \langle n_y \rangle, \langle n_z \rangle)$
\\
\hline

$z>0,\,x=y=0,$
& FM
&  $(0,0,1)$
\\
\hline

$y= z,\, x=0$
& PM
& $(0,\frac{1}{2},\frac{1}{2})$
\\
\hline

$y\gg z,\, x=0$
& AFM
& $(0,\frac{7}{9},\frac{2}{9})$
\\ 
\hline

$x\gg y,\, z=0$
& AFM*
& $(\frac{2}{3},\frac{7}{27},\frac{2}{27})$
\\
\hline

$x\gg z,\, y=0$
& HAFM
& $(\frac{2}{3},0, \frac{1}{3})$
\\
\hline

$x=y,\, z=0$
& HFM
& $(\frac{1}{2},\frac{1}{2},0)$
\\
\hline

\end{tabular}

\caption{Overview of the vertex occupations for the different vertex orders with their corresponding $(x,y,z)$. The first column are the approximate range of parameter values, the second column gives the names and the third column gives the values of $\langle \mathbf{n} \rangle$ in the limit as the system size $L\to \infty$.}
\label{tab:overviewofphases}
\end{table}

\subsection{Conventional Orders: FM, PM, AFM}
As explained in Sec.~\ref{sec:fourvertx=0}, these orders can be understood simply in terms of a superspin model for $x=0$ in which the circulation of the small triangular plaquettes defines a superspin $S_p=\pm 1$. The different vertex occupations are calculated directly via the constraints imposed by the hierarchy of $y$ and $z$. 

The FM regime reduces to a two-vertex model since $z$-type vertices can tile the Sierpinski gasket. The PM regime is a four-vertex model with $y=z$ where they appear in equal proportions for typical states. The AFM regime when $y \gg z$ is over-constrained due to the frustration in anti-aligning 3 adjoining plaquettes (as in the frustrated triangular Ising AFM). This means that for every set of 3 plaquettes on the fractal at the microscopic level, there is exactly one $z$-type vertex that cannot be removed. However, due to the finite ramification, the magnetic frustration only exists on the microscopic level of $n=1$, after that it is possible to use $D_n$ configurations, with mixed external vertex circulations, to guarantee that every added vertex is a $y$. For a gasket of level $n$, there are $3^n$ triangular plaquettes. Since there is one $z$-type vertex for every $3$ of these, we should have an average $z$-type vertex density $\langle n_z\rangle=3^{n-1}/N_v$ where $N_v$ is the total number of vertices. Hence, as $n\to\infty$, $\langle n_z\rangle=2/9$. We know $\langle n_x \rangle =0$, hence $\langle n_y \rangle=7/9$ such that  $\sum \langle n_\alpha\rangle=1$ is normalised.

\subsection{Fractal Orders:  AFM*, HAFM, HFM}
In this regime, we have $x>0$ such that there are strings of $x$-type vertices. The extended $1$D nature of the strings gives rise to phases whose behavior is dependent on the overall fractal structure.

\subsubsection{AFM*}
The AFM* order when $x\gg y\gg z$ is essentially identical to the AFM order, but is just delayed by one step in terms of the scale, $n$. This is because on the microscopic level, we maximize the number of $x$-type vertices as shown in Fig.~\ref{fig:vertexorders}(c). For the $n=1$ generation, $F_1 \gg Q_1$ since the $F_1$ configuration gives three $x$-type vertices compared to only $2$ for $Q_1$. At scales bigger than $n=1$, since $y>z$ we now seek to maximize the number of $y$-type vertices, which is an identical problem to the aforementioned AFM. Hence, this model is a true six-vertex model with non-zero occupation of all vertex types. There is effectively one $x$-type vertex per plaquette such that $\langle n_x \rangle =3^n/N_v\sim2/3$ as $n\to \infty$. Of the remaining $1/3$ of the vertices, the fractions derived earlier for the AFM give $\langle n_y \rangle=7/27$ and $\langle n_z \rangle=2/27$. When $x\gg y\gg z$, this is in excellent agreement with numerics, but deviates when $x\gg y\gg z$ differ by less than one order of magnitude.

\subsubsection{HAFM}
For the HAFM phase ($x>z,\,y=0$), there is considerable variance in the appearance of typical states, because all the states without any $y$-type vertices are degenerate in free energy at large length scales as argued in the main text. While there is an energetic tendency to anti-align the dual plaquettes as shown in Fig.~\ref{fig:vertexorders}, the sum is equally weighted over all loop configurations when coarse-grained due to the entropic contribution.

The vertex occupations always have $\langle n_y \rangle=0$ due to $y=0$ in this case, but $\langle n_x \rangle$ and $\langle n_z \rangle$ vary depending on the ratio $x/z$. When $x\approx z$, since all the allowed states are degenerate in this limit, it follows that $\langle n_x \rangle \approx \langle n_z \rangle\approx 1/2$. In the case when $x \gg z$, it is energetically favourable to maximize the loop length. Hence, the scenario is similar to the AFM*, with maximum total loop length, but where all the $y$-type vertices are instead $z$-type. This agrees with the results for the two phases in Table.~\ref{tab:overviewofphases}.

\subsubsection{HFM} \label{sec:HFM}
This regime corresponds to the point $(x,y,z)=(1,1,0)$. When $z$-type vertices are energetically forbidden ($z=0$), there are only two allowed configurations. This is explained in Sec.~\ref{sec:fourvertz=0} by there only being a single flippable loop (Euler circuit in Fig.~\ref{fig:eulercircuit}). Since flipping the loop exchanges all $x$-type and $y$-type vertices, we know that in the thermodynamic limit, $\langle n_x  \rangle =\langle n_y \rangle=1/2$ when averaged over the two configurations.

\begin{figure}
    \centering
    \includegraphics[width=\linewidth]{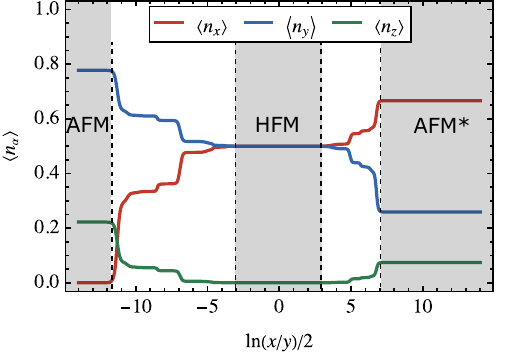}
    \caption{Cascade of crossovers in $\langle n_\alpha \rangle$ between the previously discussed regimes highlighted in gray for fixed $z=10^{-10}$ and $xyz=1$ and $n=25$ generations. For $\ln(x/y)/2\approx 3$, we see that a small gap opens between $\langle n_x \rangle$ and $\langle n_y \rangle$ accompanied by adding few energetically expensive $z$-type vertices.}
    \label{fig:cascadecrossover}
\end{figure}

The entropy in this four-vertex limit is non-extensive, unlike the other four-vertex models. Hence, it is  unstable to perturbations of the form $z\ll x,y$; although $z>0$ may have a prohibitive energy barrier, there is an extensive gain in entropy from introducing it as a defect. This instability to $z$-type defects gives rise to a cascade of crossovers (Fig.~\ref{fig:cascadecrossover}, where several different states become relevant as we tune the ratio $x/y$ for fixed $z\ll x,y$. This phenomenon does not occur for any of the other boundaries between the different regimes.

\section{Asymptotic Recursion Solutions}
\label{sec:asymrecur}
The non-linear coupled recursion relations given in the End Matter do not admit closed-form solutions. However, in certain special cases, including the four-vertex limits, it is possible to extract asymptotic solutions that capture the key behavior of the recursion ratios $R_n=\mathcal{Z}_n/Q_n$ and $P_n=D_n/F_n$.

\subsection{$x=0$: AFM vs. PM vs. FM}
\label{sec:x=0asym}
In the limit $x=0$, we find that $C_n^+=C_n^-=0$ for all $n>0$.
In this simplified regime, we then find that the partition function is given by $\mathcal{Z}_n=2F_n+6D_n$. Hence, we can characterize it with $P_n\equiv D_n/F_n$ and the recursion is given by
\begin{equation}
    P_{n+1}
=
\frac{
P_n^2 \left(2y^2 + 4yz + 3z^2\right)
+ P_n \left(y^2 + 4yz + z^2\right)
+ y^2
}{
P_n^2 \left(y^2 + 4yz + 4z^2\right)
+ P_n \left(3y^2 + 4yz - z^2\right)
+ z^2
},
\label{eqn:Pnyz}
\end{equation}
where the initial value $P_0=0$ since $D_0=0$ always. The fixed point equation from solving $P_{n+1}=P_n=P_*$ is
\begin{equation}
    (P_*-1)(P_*(y+2z)+y)^2=0
\end{equation}
with solutions $P_*=1$ (stable) and $P_*=-y/(y+2z)$ (unstable) for any $y,z\geq 0$.
For $z\to 0$, the recursion initially blows up with $P_1= y^2/z^2$. However, it settles rapidly to $P_*=1$ as we see in the projected RG flow figure in the main text. For $y=z$, $P_{n}=1$ for all $n>0$ which is the completely uncorrelated PM. If $y=0$, we have the FM phase (only $z$ non-zero) which acts as an unstable fixed point $P_*=0$.
Hence, in agreement with the Ising model on the Sierpinski gasket, we see that our superspin picture recovers the expected lack of ordering at finite temperature, since we flow to the paramagnet $P_*=1$ for any finite temperature. 

The flow away from the unstable fixed point for the FM when $y>0$ can be calculated by taking $y/z \ll 1$, and using Eq.~\ref{eqn:Pnyz} to show that
\begin{equation}
    P_{n+1}-P_{n}= \bigg(\frac{y}{z}\bigg)^2 +\mathcal{O}\bigg(y^3z^{-3}\bigg).
\end{equation}
Hence, given $P_0=0$, we have $P_n\sim n (y/z)^2$ for $y \ll z$ from neglecting all but the lowest order term. The crossover length scale, $L_c$, is decided when the FM solution assuming $y=0$ breaks down and $P_n\sim 1$. Hence, $n_c\sim (z/y)^2$ such that the length scale $L_c=2^{n_c}$ is given by $L_c\sim2^{z^2y^{-2}}$, which diverges rapidly as $y\to0$. We take $n=30$ for the projected RG flow in the main text, which would imply that for $y/z > 1/\sqrt{30}\approx 0.18$ it will not yet have converged to the final value $P_*$.

\subsection{$y=0$: HAFM vs. FM}
\label{sec:HAFM}
The recursion relations for $y=0$ are given by
\begin{equation}
\begin{aligned}
    F_{n+1}^+&=z^3(F_n^+)^3+x^3 (C^+_n)^3,\\ F_{n+1}^-&=z^3(F_n^-)^3+x^3 (C^+_n)^3,\\ C_{n+1}^+&=xz^2 (F_n^-)(C_n^+)^2+x^2z(C_n^+)^3,
\end{aligned} 
\label{eqn:yforbidrecurse}
\end{equation}
such that 
\begin{equation}
    R_{n+1}=\frac{\mathcal{Z}_{n+1} }{Q_{n+1}}=\frac{F_{n+1}^++F_{n+1}^-}{C_{n+1}^+}=\frac{zR_n^2}{2x}-R_n+\frac{2x}{z}.
\end{equation}

We first rescale these equations to simplify the analysis, then quote the results in terms of the original parameters. Defining $R_*=2x/z$, and rescaling $\rho_n=R_n/R_*$, we find that the recursion relation is
\begin{equation}
    \rho_{n+1}=\rho_n^2-\rho_n+1.
    \label{eqn:recurserescale}
\end{equation}
The difference between consecutive values is given by
\begin{equation}
    \rho_{n+1}-\rho_n=(\rho_n-1)^2.
    \label{eqn:rho_increment}
\end{equation}
From this, we can directly read off that there is a unique fixed point $\rho_*=1$ of algebraic multiplicity two. Since the sign of Eq.~\ref{eqn:rho_increment} is always positive, the fixed point is semi-stable. 

There are two key regimes to consider. If $\rho_0>1$, then we find that $\rho_n \to \infty$. Using Eq.~\ref{eqn:recurserescale}, we find that $\rho_{n+1}\sim\rho_n^2$ such that $\rho_n\sim \gamma(x/z)^{2^n}$, where the function $\gamma(t)$ is independent of $n$, but depends on the ratio $t=x/z$. If $0<\rho_0<1$, the positive increment causes it to approach $\rho_n \to \rho_*$ from below. To determine the rate of  convergence and find the smallest correction in $n$, we define 
\begin{equation}
    v_n=1/(\rho_*-\rho_n)
    \label{eqn:vndef}
\end{equation} such that $v_n\to \infty$ as $\rho_n\to\rho_*$. Substituting $v_n$ into Eq.~\ref{eqn:rho_increment}, we find that
\begin{equation}
    v_{n+1}-v_n=1+\frac{1}{v_n-1}.
    \label{eqn:vnrecur}
\end{equation}
We know that $v_n\to \infty$, hence we can neglect the latter term to find that $v_{n+1}-v_n\sim1$ such that $v_n\sim n$. Putting this crude result into Eq.~\ref{eqn:vnrecur} for $v_n$ on the RHS, we find the next order approximation is
\begin{equation}
    v_n-v_0=\sum_{k=0}^{n-1} (v_{k+1}-v_k)=\sum_{k=0}^{n-1}\bigg(1+\frac{1}{k}+\mathcal{O}(1/k^2)\bigg).
\end{equation}
Summing the harmonic series, we find that
\begin{equation}
    v_n\sim n+\ln(n).
\end{equation}
Substituting this back into Eq.~\ref{eqn:vndef} to find $\rho_n$, we see that
\begin{equation}
    \rho_n=\rho_*-\frac{1}{v_n}=1-\frac{1}{n}+\mathcal{O}(\ln(n)/n^2),
\end{equation}
where there are sub-leading corrections associated with the logarithm. 

Converting this back to the variable $R_n$, we find that
\begin{equation}
R_{n}\sim\begin{cases}
\frac{2x}{z}\bigg(1-\frac{1}{n}+\mathcal{O}\bigg(\frac{\log(n)}{n^2}\bigg)\bigg), & x>z \\
 \gamma(x/z)^{2^n}, & x<z \\
2, & x=z,
\end{cases}
\label{eqn:Rny=0}
\end{equation}
for a function $\gamma(t)>1$ for $0<t<1$ which depends on the ratio $x/z$ determined by the asymptotics. We see that there is a crossover between the HAFM phase with $R_*< \infty$ and the FM phase exactly when $x=z$.

If these results are phrased in terms of the system size $L=2^n$, we see that the finite-size convergence to the true fractal thermodynamic limit $n\to\infty$ is like $1/\log_2(L)$ for $x>z$ as claimed in the main text.


\subsection{$z=0$: HFM}
\label{sec:HFMasym}
When $z=0$, we find that $D_n=0$ for all $n$. Hence, the relevant recursion relations are
\begin{align}
    F_{n+1}&=x^3\big((C^+_{n})^3+(C_n^-)^3\big),\\
    C_{n+1}^+&=xy^2(C_n^{-})^2F_n^+,\\
    C_{n+1}^-&=xy^2(C_n^+)^2F_n^-.
\end{align}
These recursion relations admit an exact solution. This result follows from the observation that since $C_0^-=0$, one of either $C_n^+$ or $C_n^-$ is always zero, allowing us to use $Q_n=C_n^++C_n^-$. Hence, the recursion relations can be simplified to 
\begin{equation}
F_{n+1}=x^3Q_{n}^3,\quad Q_{n+1}=xy^2Q_n^2F_n.
\end{equation}
The exact solution to these recursion relations with initial condition $F_0=1$ and $Q_0=1$ is given by
\begin{equation}
    R_{n+1}=\frac{2F_{n+1}}{Q_{n+1}}=\frac{2x^2Q_n}{y^2F_n}=\frac{4x^2}{y^2R_{n}},
\end{equation}
which holds for all $x,y$. Given $R_0=2$, the solution is
\begin{equation}
R_n =
\begin{cases}
2, & n \ \text{even}, \\[4pt]
2\left(\dfrac{x}{y}\right)^2, & n \ \text{odd}.
\end{cases}
\label{eqn:RnHFM}
\end{equation}
Interestingly, this does not flow to a fixed value, but oscillates unless $x=y$. We can then use the solution for $R_n$ to find the form of $F_n$ and $Q_n$. We find that
\begin{equation}
    F_{n+1}=\begin{cases}
        x^3F_{n}^3,\, n \text{ even},\\
        \frac{y^6}{x^3}F_n^3,\, n \text{ odd}.
    \end{cases}
\end{equation}
Since $F_0=1$, we see that we accumulate powers of $x$ in a nested series which is of the form,
\begin{equation}
    N_{x,n}=3^n-3^{n-1}+...+(-1)^{n-1} 3=\frac{3}{4}(3^n-(-1)^n).
\end{equation}
We know that the total number of bulk vertices $N_v=3(3^n-1)/2$. Since at generation $n$, $N_{z}(n)=0$ (since $z=0$), we have that
\begin{equation}
    N_{y}(n)=\frac{3}{4}(3^n+(-1)^n-2).
\end{equation}
Substituting these powers, we find that 
\begin{align}
        F_n&=\bigg( x^{3^n-(-1)^n} y^{3^n+(-1)^n-2}\bigg)^{3/4},\\
        Q_n&=\bigg( x^{3^{n+1}+(-1)^n-4} y^{3^{n+1}-(-1)^n-2}\bigg)^{1/4}.
\end{align}
In the limit as $n\to \infty$, we see that the number of $x$ and $y$ vertices in the bulk is equal for $z=0$ and any $x,y$. We note however that the coefficients of $F_n$ and $Q_n$ do not grow with size, such that the entropy in this phase is given by $S_{z=0}=\ln(2)$, just like the FM. This is where the name HFM comes from, since the states also have a hierarchical structure. This small entropy means that by adding an expensive energy defect when $z>0$, we can still reduce the free energy through the entropic gain of being able to place the defect. As mentioned earlier, this is the origin of the cascade of crossovers in Fig.~\ref{fig:cascadecrossover}.

\subsection{$y=z\equiv \epsilon$: Ice regime vs. PM}

\subsubsection{$R_n$ for general $x$}\label{sec:RnIceRegime}
\begin{figure}
    \centering
    \includegraphics[width=\linewidth]{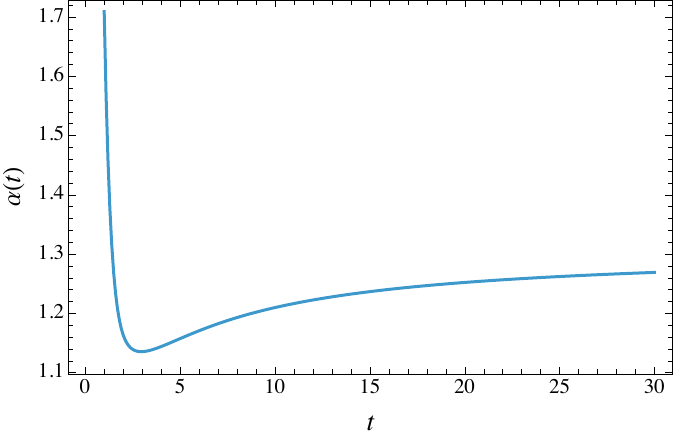}
    \caption{Plot of the parameter $\alpha(t)$ (Eq.~\ref{eqn:Rnalphat}) as a function of $t$. $\underset{t\to0}{\lim}\alpha(t)\to\infty$ is cut out of the plot.}
    \label{fig:alphat}
\end{figure}

This limit admits a reduced set of recursion relations only in terms of $\mathcal{Z}_n$ and $\mathcal{Q}_n$. Using $y=z=\epsilon$, these reduced relations are given by 
\begin{equation}
    \mathcal{Z}_{n+1} =\epsilon^3\mathcal{Z}_n^3+2x^3Q_n^3, \quad Q_{n+1}=\epsilon x^2Q_{n}^3 + \epsilon^2 x Q_n^2\mathcal{Z}_n.
\end{equation}
These equations can be asymptotically solved by studying the ratio of the $R_n=\mathcal{Z}_n/Q_n$ given by
\begin{equation}
    R_{n+1}=\frac{\epsilon^3 R_n^3+2x^3}{\epsilon^2 xR_n+\epsilon x^2},
    \label{eqn:y=zRrecur}
\end{equation}
where in the final step we used $x\epsilon^2=1$ from our normalization choice $xyz=1$. Studying the recursion relation in Eq.~\ref{eqn:y=zRrecur}, it has a single unstable fixed point at $R_*=-1.2x/\epsilon$. Given that our initial condition is $R_0=2$, this means all solutions flow to infinity $R_n \to \infty$ with asymptotic solution given by
\begin{equation}
    R_n\sim \bigg(\frac{x}{\epsilon}\bigg)\bigg(\alpha\bigg(\frac{x}{\epsilon}\bigg)\bigg)^{2^n}.
    \label{eqn:Rnalphat}
\end{equation}
Crucially, the function $\alpha(t)>1$ for all $t$, which ensures the divergence of $R_n$. The form of the function is shown in Fig.~\ref{fig:alphat} where $\alpha(1)=\alpha\approx1.709$. 

\subsubsection{$P_n$ for $x=y=z=1$}\label{sec:PnIceRegime}
As mentioned in Sec.~\ref{sec:x=0asym}, the paramagnetic parameter $P_n$ has two values, $P_*=1$ which is stable and $P_*=0$ which is unstable with respect to perturbations $y\neq0$. We want to understand the asymptotics of $P_n$ to the stable solution $P_*=1$ for the ice point ($x=y=z=1$).

Consider the quantity
\begin{equation}
   1 - P_n=
\frac{F_n - D_n}{F_n}.
\label{eqn:1-PnstepA}
\end{equation}
As we approach $P_n=1$, we know that $F_n \simeq D_n$ such that we can substitute $\mathcal{Z}_n=2(F_n+3D_n)\simeq8F_n$ up to corrections which decay at least exponentially in $n$. Additionally, using the full recursion relations for $F_n$ and $D_n$ quoted in the End Matter, we find that for the special case of the ice point that
\begin{align}
    F_n-D_n&=\Bigl(
(C_{n-1}^{-})^2 - (C_{n-1}^{+})^2
\Bigr)
\Bigl(
C_{n-1}^{-} - C_{n-1}^{+}
\Bigr)\\
&=\Bigl(
C_{n-1}^{-} + C_{n-1}^{+}
\Bigr)\Bigl(
C_{n-1}^{-} - C_{n-1}^{+}
\Bigr)^2\\
&=Q_{n-1}\Bigl(
C_{n-1}^{-} - C_{n-1}^{+}
\Bigr)^2.
\end{align}
Combining these results together with Eq.~\ref{eqn:1-PnstepA} and the definition $R_n=\mathcal{Z}_n/Q_n$ we find that
\begin{equation}
1-P_n
\simeq
\frac{8 Q_{n-1}
\left(
C_{n-1}^{-} - C_{n-1}^{+}
\right)^2}{\mathcal{Z}_n}.
\label{eqn:1minusPnstepB}
\end{equation}

We have already found the asymptotic solution $R_n\sim \alpha^{2^n}$ and $\mathcal{Z}_n\sim \Omega_0^{3^n}$ in the main text. Now we need to solve for $Q_n$ and $C_{n-1}^+-C_{n-1}^-$.

Substituting in the definition of $R_n$, we find that $Q_n=\mathcal{Z}_n/R_n\sim \Omega_0^{3^n}/\alpha^{2^n}$. To solve for the difference $C_{n-1}^+-C_{n-1}^-$, we need to write down the recursion relation. Using the full equations from the End Matter once again,
\begin{align}
    C_{n+1}^--C_{n+1}^+&=(C_n^--C_n^+)Q_n^2\\
    &=(C_n^--C_n^+) \Omega_0^{2\cdot 3^n}\alpha^{-2\cdot 2^n},
    \label{eqn:∆Crecurse}
\end{align}
where we substituted in the asymptotic solution for $Q_n$. In the asymptotic limit for $Q_n$, the recursion relation in Eq.~\ref{eqn:∆Crecurse} has a solution given by
\begin{equation}
    C_n^--C_n^+\sim-C_0^+ \frac{\alpha^2}{\Omega_0}\Omega_0^{3^n}\alpha^{-2\cdot2^n}.
\end{equation}
Finally, substituting in all the asymptotic forms into Eq.~\ref{eqn:1minusPnstepB}, we find that
\begin{align}
    1-P_n&\sim 8 \frac{Q_{n-1}}{\mathcal{Z}_n} \left(
C_{n-1}^{-} - C_{n-1}^{+}
\right)^2\\
&\sim\frac{8}{\Omega_0^{3^n}} \frac{\Omega_0^{3^{n-1}}}{\alpha^{2^{n-1}}}\bigg( C_0^+ \frac{\alpha^2}{\Omega_0}\Omega_0^{3^{n-1}} \alpha^{-2^n}\bigg)^2\\
&\sim8 \bigg(\frac{\alpha^2}{\Omega_0} \bigg)^2 \Omega_0^{3^n(-1+1/3+2/3)}\alpha^{-2^n(2+1/2)}\\
&\sim8 \bigg(\frac{\alpha^2}{\Omega_0} \bigg)^2 \alpha^{-\frac{5}{2}\cdot2^n}.
\end{align}
This gives the key result quoted in the main text that $1-P_n$ decays super-exponentially in the depth of the fractal, $n$, and hence decays exponentially in the length $L=2^n$.

\section{String Tension in the Bulk}
\label{sec:stringtensionbulk}

We derive the string tension for one of the charges placed in the bulk to examine its effect in the thermodynamic limit. We find that in the confining case, the bulk is irrelevant, since string contributions that take long meandering trajectories into the bulk are suppressed exponentially in their length. By contrast, for the HAFM case, these meandering strings must be explicitly included, since all trajectories have equal free energy at large scales, but they cancel to give the same tensionless result as when they are only on the corners.

\subsection{Ice Point $(x=y=z)$}
The full calculation of every possible string trajectory in the bulk reduces to the simple string tension calculation for charges on the corners presented in the main paper with corrections like negative powers of $R_n$. 

\begin{figure*}
    \centering
    \includegraphics[width=\linewidth]{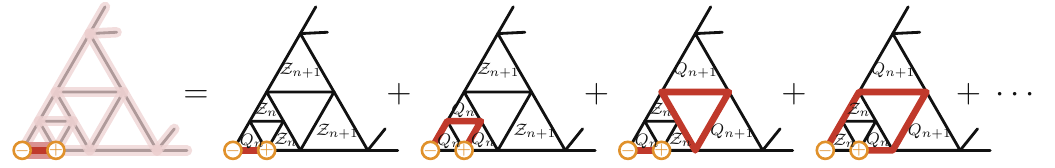}
    \caption{Diagrams corresponding to the meandering string path between a pair of fixed charges (orange) for the ice point $x=y=z=1$. The diagram on the left side represents the total weight associated with the configuration where the two charges are fixed, and the sum on the right are some of the contributions. The weight of a given contribution is the product of its plaquette and vertex factors, and multiplied by $2$ for every closed loop. Since $x=y=z=1$ for the ice point, the third diagram has weight $W_3=2Q_n\mathcal{Z}_n^2Q_{n+1}^2\cdots$, where the ellipsis indicates the other bulk plaquette configurations that have not been specified.}
    \label{fig:meanderstring}
\end{figure*}

To handle the case where a charge is in the bulk, we introduce new recursion variables which have a constraint of a single charge placed inside of the bulk at the $n$th generation (example in Fig.~\ref{fig:meanderstring}). We denote these with a superscript $n$ such that the initial values at a certain scale are given by
\begin{equation}
    \begin{aligned}
        \mathcal{Z}^{(n)}_{n+1}&=Q_n\mathcal{Z}_n^2+Q_n^3=Q_n\mathcal{Z}_n^2(1+R_n^{-2}),\\
    Q^{(n)}_{n+1}&=2Q_n^3=2\mathcal{Z}_n^3R_n^{-3},
    \end{aligned}
    \label{eqn:IC_constrain}
\end{equation}
where we have used the diagrammatics to get all the recursion terms, and used the definition $R_n=\mathcal{Z}_n/Q_n$. This is a useful way of writing the equations, because $R_n \to \infty$ as $n\to\infty$ means that negative powers of  $R_n$ can be understood as perturbative corrections. $R_n^{-1}$ is always raised to larger powers than those in Eq.~\ref{eqn:IC_constrain} as we iterate the bulk to be larger. Hence, we have already captured the lowest-order part of the correction with this first term. We find that
\begin{equation}
    \mathcal{Z}^{(n)}_{n+m}=Q_n\mathcal{Z}_n^2 \mathcal{Z}_{n+1}^2\cdots \mathcal{Z}_{n+m-1}^2\big(1+R_n^{-2}+\mathcal{O}(R_n^{-3}) \big).
    \label{eqn:Znm}
\end{equation}
We calculate a few diagrams explicitly in Fig.~\ref{fig:meanderstring} to show the terms of Eq.~\ref{eqn:Znm}. We want to prove $\mathcal{Z}^{(n)}_{n+m}/\mathcal{Z}_{n+m}\approx Q_n/\mathcal{Z}_n$ up to small corrections to support our claim that the bulk calculation is equivalent to that for just the corners. $\mathcal{Z}^{(n)}_{n+m}$ counts the weight of all the states with the charge embedded in the bulk, whereas $\mathcal{Z}_n^{(n)}=Q_n$ assumes the charges are placed on the external corners. 
Writing $\mathcal{Z}_{n+2}$ in a similar way using the recursion relations, we find that
\begin{equation}
    \mathcal{Z}_{n+2}=\mathcal{Z}_n^3\mathcal{Z}_{n+1}^2\big(1+2R_n^{-3}+(R_n^{-2}+R_n^{-3})R_{n+1}^{-2}\big),
\end{equation}
where we see that the smallest loop correction we can make is $2R_n^{-3}$, which holds even as we extend the bulk with $\mathcal{Z}_{n+m}$. Hence, the ratio for arbitrary sizes goes like
\begin{equation}
    \begin{aligned}
    \frac{\mathcal{Z}^{(n)}_{n+m}}{\mathcal{Z}_{n+m}}&=\frac{Q_n}{\mathcal{Z}_n} \frac{(1+R_n^{-2}+\cdots)}{(1+2R_n^{-3}+\cdots)}\\
    &=\frac{Q_n}{\mathcal{Z}_n}\big(1+R_n^{-2}+\mathcal{O}(R_n^{-3})\big).
    \end{aligned}
\end{equation}
In support of the claims of the main text, we see that up to corrections in $R_n$ that go to zero as $n\to\infty$, the bulk behaviour is the same as the calculation for the corners, such that a good approximation is just to take the shortest string.

\subsection{HAFM $(x>z,\, y=0)$}
For the HAFM, we have a deconfinement and will hence have vanishing string tension. This means that we cannot apply the same argument of neglecting large string trajectories as for the ice point ($x=y=z=1$) where the theory was confining. Below we use the result from Sec.~\ref{sec:HAFM} for the recursion relations when $y=0$.

Applying the same procedure as for the ice regime, we calculate the constrained versions $F^{-,(n)}_{n+m}$ and $C^{+,(n)}_{n+m}$ first starting with their initial values for $m=1$,
\begin{equation}
    \begin{aligned}
        F^{-,(n)}_{n+1}&=x^2(C_n^+)^3+z^2(C_n^+)(F_n^-)^2\\
        C_{n+1}^{+,(n)}&=2xz(C_n^+)^2. 
    \end{aligned}
    \label{eqn:IC_constrainy=0}
\end{equation}
Subsequent values are calculated recursively from a modified Eq.~\ref{eqn:yforbidrecurse} where we replace one of the variables with the constrained version in each term:
\begin{equation}
\begin{aligned}
    F_{n+m+1}^{-,(n)}&=z^3F_{n+m}^{-,(n)}(F_{n+m}^-)^2+x^3 C_{n+m}^{+,(n)}(C^+_{n+m})^2,\\ 
    C_{n+m+1}^{+,(n)}&=xz^2 F_{n+m}^{-,(n)}(C_{n+m}^+)^2+x^2zC_{n+m}^{+,(n)}(C_{n+m}^+)^2.
\end{aligned} 
\label{eqn:yforbidrecurse_constrain}
\end{equation}
The tensionless behavior arises in the limit where the asymptotic solution of Sec.~\ref{sec:HAFM} where $F_n^-/C_n^+=x/z$. Substituting in the asymptotic relation, we get simplified initial conditions,
\begin{equation}
    \begin{aligned}
        F^{-,(n)}_{n+1}&=2x^2(C_n^+)^3,\\
        C_{n+1}^{+,(n)}&=2xz(C_n^+)^3, 
    \end{aligned}
    \label{eqn:IC_constrainy=0_asymp}
\end{equation}
and simplified recursion relations,
\begin{equation}
\begin{aligned}
    F_{n+m+1}^{-,(n)}&=x^2(C^+_{n+m})^2\big(zF_{n+m}^{-,(n)}+xC_{n+m}^{+,(n)}\big),\\ 
    C_{n+m+1}^{+,(n)}&=xz(C_{n+m}^+)^2(z F_{n+m}^{-,(n)}+xC_{n+m}^{+,(n)}).
\end{aligned} 
\label{eqn:yforbidrecurse_constrain_asmy1}
\end{equation}
These recursion relations keep the ratio $F_{n+m+1}^{-,(n)}/C^{+,(n)}_{n+m+1}=x/z$ fixed for all $m$ if $n$ is sufficiently large that the asymptotic results hold, which agrees with Eq.~\ref{eqn:IC_constrainy=0_asymp}. Hence, we can make a final simplification of the recursion relations to
\begin{equation}
\begin{aligned}
    F_{n+m+1}^{-,(n)}&=2x^3(C^+_{n+m})^2C_{n+m}^{+,(n)},\\ 
    C_{n+m+1}^{+,(n)}&=2x^2z(C_{n+m}^+)^2C_{n+m}^{+,(n)}.
\end{aligned} 
\label{eqn:yforbidrecurse_constrain_asym2}
\end{equation}
We define the string tension from the difference in free energy under extending the string but sending $m\to\infty$ such that the charge at generation $n$ is embedded deep into the bulk. The difference in free energies is
\begin{equation}
    \Delta f=f(l)-f(l/2)=\lim_{m\to\infty} \log \bigg(\frac{F^{-,(n)}_{n+m}}{F^{-,(n+1)}_{n+m}} \bigg),
\end{equation}
where $l=2^n$ is the larger length of the string. For the first recursion step, we find that $F_{n+m}^{-,(n)}=2x^3 (C_{n+m-1})^2C^{+,(n)}_{n+m-1}$ from Eq.~\ref{eqn:yforbidrecurse_constrain_asym2}. Thus, canceling factors, we find that
\begin{equation}
    \Delta f=\lim_{m\to\infty} \log \bigg(\frac{C^{+,(n)}_{n+m-1}}{C^{+,(n+1)}_{n+m-1}} \bigg).
\end{equation}
We can then subsequently repeat this process $m$ times until we reach the initial condition, $C^{(n+1)}_{n+2}=2xz(C_{n+1}^+)^3$ from Eq.~\ref{eqn:IC_constrainy=0_asymp}. Taking the numerator one generation further with the recursion, we find that
\begin{equation}
    \Delta f=\lim_{m\to \infty} \log \bigg( \frac{2x^2z (C_{n+1}^+)^2(C_{n+1}^{+,(n)})}{2xz (C_{n+1}^+)^3 }\bigg).
\end{equation}
Again canceling factors, and using Eq.~\ref{eqn:IC_constrainy=0_asymp}, we find that
\begin{equation}
    \Delta f=\lim_{m\to \infty} \log \bigg( x\frac{2xz(C_n^+)^3}{C_{n+1}^+}\bigg)=0,
\end{equation}
where the final equality with zero comes from the asymptotic recursion $C_{n+1}^+=2x^2z(C_n^+)^3$. Hence, we see the asymptotic limit supports the claim in the main text that the charges are joined by a tensionless string in the HAFM regime where $y=0$ and $x>z$.

\section{Experimental Implementation}
\label{sec:ASIimplementation}
A well-established platform for probing the collective physics of interacting magnets is artificial spin ice (ASI). This platform consists of ``islands" made of ferromagnetic material with effectively two different possible magnetization directions which can be modeled as  Ising variables $\sigma_j$. By arranging arrays of these islands, one can engineer the interactions such that spin-ice constraints are enforced in the low-energy states. 

The challenge of ASI is that simply placing the islands on the edges of the desired graph geometry does not generally give rise to the desired degenerate ice point weights, except in the notable exception of kagome ASI where it is guaranteed by symmetry \cite{KagomeASI1, Moller2009MultipoleASI}. It has been noted that modifying square ASI to rectangular ASI can be used to  generate degeneracy \cite{Ribeiro2017RectangularASI, Nascimento2024GaugeFieldPlanarASI}. Here, we explore a generalized technique for engineering degeneracy of different vertex configurations by adjusting the positions of the tips of the islands. This yields a high degree of tunability of relative vertex weights, including the construction of a family of  parameters with vertex degeneracy, and an approximation of the parameters where fractal deconfinement takes place. 

The energy of different vertex configurations is calculated within the dumbbell model. In this approximation, we model the magnetic islands of length $l$ with two magnetic charges of opposite sign placed at on the tips at either end with values given by the dipole moment, $\mu$, such that they have magnetic charges $q_m=\mu/l$ \cite{Castelnovo2008}. This approximation works well when higher order terms in the multipole expansion can be neglected, which is true when $w\ll a_j \ll l$, where $w$ is the width of the islands and $a_j$ is the distance of the $j$th tip from the vertex.

\begin{figure}
    \centering
    \includegraphics[width=\linewidth]{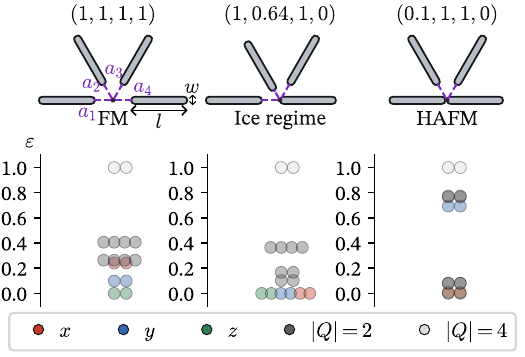}
    \caption{Examples of different vertex configurations specified by the radial distances of the island tips from the vertex $(a_1,a_2,a_3,a_4)$. $a_j$ have been tuned to different Boltzmann weights $(x,y,z)$ as measured by the scaled dimensionless energy $\epsilon=(E-E_\text{min})/E_\text{max}$. The white points represent the $Q=\pm 4$ states which are always highest in energy $\epsilon=1$. $(1,1,1,1)$, no vertex tuning, supports the FM with Boltzmann weights $z>y>x$. $(1,0.64,1,0)$ is tuned to approximate the ice point with $x\approx y\approx z$. $(0.1,1,1,0)$ is tuned to approximate the edge of the fractal regime with $x\approx z \gg y$. All calculations performed by placing charges on the tips of islands and calculating the Coulomb energy.}
    \label{fig:Dumbbell}
\end{figure}

We study the energy of vertex configurations in terms of four parameters $(a_1,a_2,a_3,a_4)$ which are the distances of the tips from the vertex as shown in purple in Fig.~\ref{fig:Dumbbell}. These rays on which $a_j,a_{j+1}$ lie are separated by an angle of $60^{\circ}$. Given these constraints, and setting $q_m^2/a=1$ for convenience as a natural energy scale, the energy of a general configuration is given by
\begin{equation}
    E_\alpha= \sum_{i> j}\frac{q_i^{(\alpha)}q_j^{(\alpha)}}{r_{ij}},
\end{equation}
where $q_i^{(\alpha)}=\pm 1$ depending on the vertex configuration.
As an example, for the $x$ configuration, we can write
\begin{equation}
\begin{aligned}
        E_x&=\frac{1}{r_{12}}+\frac{1}{r_{34}}-\bigg(\frac{1}{r_{23}}+\frac{1}{r_{24}}+\frac{1}{r_{13}}+\frac{1}{r_{14}} \bigg)\\
        &= -\sum_{i<j} \frac{1}{r_{ij}}+2\bigg(\frac{1}{r_{12}}+\frac{1}{r_{34}}\bigg).
\end{aligned}
\label{eqn:ExDumbbell}
\end{equation}
Under permutation symmetry of labels $1,2,3,4$, we can write similar equations for $y,z$. 

Aside from the local geometry of each vertex, when joining these together to form the full fractal, they generally force islands to have different lengths, hence different total magnetic moments. However, in the limit where the dumbbell model applies, $w\ll a_j \ll l$, effects due to different island lengths are also negligible--what is important at any rate is the magnetisation per unit length of the island, which is naturally approximately independent of the island length for, say, a permalloy island.

\begin{table}[h]
\centering
\small
\renewcommand{\arraystretch}{1.25}
\setlength{\tabcolsep}{4pt}

\begin{tabular}{p{0.10\linewidth} p{0.25\linewidth} p{0.25\linewidth} p{0.25\linewidth}}
\hline
$n$
& $(1.0,0.2,0.6)$
& $(0.9,0.1,1.0)$
& $(1.0,0.8, 0.9)$
\\
\hline

$1$
& $2.60$
& $2.47$
& $2.38$
\\
\hline

$2$
& $2.53$
& $2.47$
& $2.29$
\\
\hline

$3$
& $2.50$
& $2.46$
& $2.29$
\\
\hline

$4$
& $2.45$
& $2.46$
& $2.17$
\\
\hline

$5$
& $2.38$
& $2.45$
& $2.04$
\\
\hline

\end{tabular}

\caption{Approximate values of $2^{d_l}$ for points $(x,y,z)$ approximating the fractal regime (first two columns) and ice regime (last column). The fractal regime is strictly only for $y=0,\,x\geq z$, and so the parameters used here have $2^{d_l}=2$ as $n\to\infty$. Although the approximation is crude, the $2^{d_l}$ values still agree well with the expected values of $2^{d_l}=5/2$, and are distinguishable from strongly confining behavior in the ice regime.}
\label{tab:approximate-alpha-values}
\end{table}
\subsection{Ice regime: $x\approx y\approx z$}
The degeneracy condition is that $E_x=E_y=E_z=E_0$, with corresponding Boltzmann factors are $\alpha=\exp(-\beta E_\alpha)$ where $\alpha \in \left\{x,y,z\right\}$. Using Eq.~\ref{eqn:ExDumbbell}, and equating it to the permuted versions for $E_y$ and $E_z$, we find that the degeneracy condition is
\begin{equation}
    \frac{1}{r_{12}}+\frac{1}{r_{34}}=\frac{1}{r_{14}}+\frac{1}{r_{23}}=\frac{1}{r_{13}}+\frac{1}{r_{24}}.
    \label{eqn:degenconstraint}
\end{equation}

In total, we have four parameters $a_j$. One of these corresponds to an overall arbitrary length scale of the problem, which we can fix to $1$. Without loss of generality, we take $a_2=1$. Eq.~\ref{eqn:degenconstraint} implements two further constraints $E_x=E_y$ and $E_y=E_z$. Hence, there is not just a single isolated solution, but in fact one continuous tuning parameter for the different configurations with degeneracy $x=y=z$. This parameter can be fixed for experimental convenience. For  example, the  solution shown in Fig.~\ref{fig:Dumbbell} is one choice from this family where the values of $a_j$ lead to well-spaced tips.

\subsection{Fractal Regime: $x\geq z\gg y$}
The vertex engineering required to observe the deconfined fractal physics is simpler than the careful matching of the degenerate case. By simply placing the tips of islands $1$ and $4$ close together with $a_1\approx a_4\approx \epsilon$, $y$-type vertices are strongly energetically disfavored since they always have like charges on tips $1$ and $4$ (Fig.~\ref{fig:Dumbbell}). 

Although the $y=0,\,x>z$ limit is not physically accessible with the geometric constraints, we show that the  approximation $x\approx z> y$ still yields an easily discernible difference in the fractal dimension of the string compared to the confined ice regime. In Table.~\ref{tab:approximate-alpha-values} we show the values of $2^{d_l}$ for some $(x,y,z)$ which are intended to be associated with different regimes. Despite all the parameter choices in Table.~\ref{tab:approximate-alpha-values} tending to $2^{d_l}=2$ as $n\to\infty$, we see that in the fractal regime around $x\geq z,y=0$, we approximately recover the fractal scaling for a finite number of generations.

\bibliography{apssamp}